# TDM/WDM over AWGR Based Passive Optical Network Data Centre Architecture

**Azza E. A. Eltraify, Mohamed O. I. Musa and Jaafar M.H. Elmirghani**
*School of Electronic and Electrical Engineering, University of Leeds, Leeds, LS2 9JT, UK*
*E-mail: {scaeae, m.musa,j.m.h.elmirghani}@leeds.ac.uk*

**ABSTRACT**
Passive Optical Data Centre Networks have been developed due to the performance limitations in current data centres to provide high performance within data centre networks. An AWGR based passive optical network data centre architecture is evaluated using a TDM/WDM multiple access technique to provision the flow of traffic among the network efficiently. A Mixed Integer Linear Programming model is developed to optimise resource allocation in the architecture. Using WDM-TDM as a multiple access technique helps in solving issues such as oversubscription and congestion by allowing servers to make simultaneous transmissions of packets in different time slots. The results show that the provisioning / allocation of resource within the architecture is improved with improvements of up to 75% in resource utilisation.

## 1. INTRODUCTION

Studies have been conducted in optimising power efficiency and architectures in data centres and core networks to be able to fulfil the rapidly increasing demand for data rates and energy efficiency [1]-[10]. Several challenges and issues have risen with conventional data centres which resulted in several researches to develop alternative designs and approaches to provide solutions for scalability, reliability and efficiency in current data centres [11]-[18], [22]-[41].

The need for more power efficient architectures has emerged from the high power consumption in data centres, where many power hungry devices are being used instead of passive devices, while many of these resources are underutilised resulting in inefficiency in resource allocation [19]. Hence, the use of PON technologies in data centres has emerged, as they have been proven to have better performance in access networks providing low cost, high capacity, low latency, scalability, and high energy efficiency. Several passive devices such as Arrayed Waveguide Grating Routers (AWGR), Fibre Bragg gratings (FBG), and star couplers/splitters have been used to implement and develop new PON architectures [20].

This paper introduces the use of WDM/TDM technique using AWGR based passive optical network data centre architectures. The use of WDM/TDM multiple access techniques is introduced to provision traffic in inter-rack, intra-rack and OLT/ONU interconnection. A mathematical optimisation model is developed to optimise the architecture. The results are compared to the results of a similar optimised architecture that uses WDM approaches only. The main purpose of the paper is to produce a more efficient architecture by dividing the available blocks (wavelengths) into smaller granularities, dividing them into several time slots.

The paper is organized as follows, Section 2 discusses the WDM-TDM PON based data centre architecture, Section 3 discusses the optimisation model, Section 4 discusses the results and finally the paper is summarised in giving conclusions in Section 5.

## 2. WDM-TDM PON-BASED DATA CENTRE ARCHITECTURE

In this architecture each PON Cell contains two AWGRs as shown in Figure.1, to provision full interconnection between the 4 racks. Each rack contains 32 servers, using 4 different wavelengths. A 1: N AWGR connects the racks and the OLT. The network interface card of each server contains an array of photo detectors for wavelength detection and tuneable lasers for wavelength selection [11].

Inter-rack communication is done through the direct AWGR link or through the OLT. The AWGR selects the wavelength for transmission according to the location of the destination server, and servers communicate according to an AWGR wavelength routing interconnection map. The availability of alternative routes provides multi-path routing and enhances load balancing at high traffic load [11].

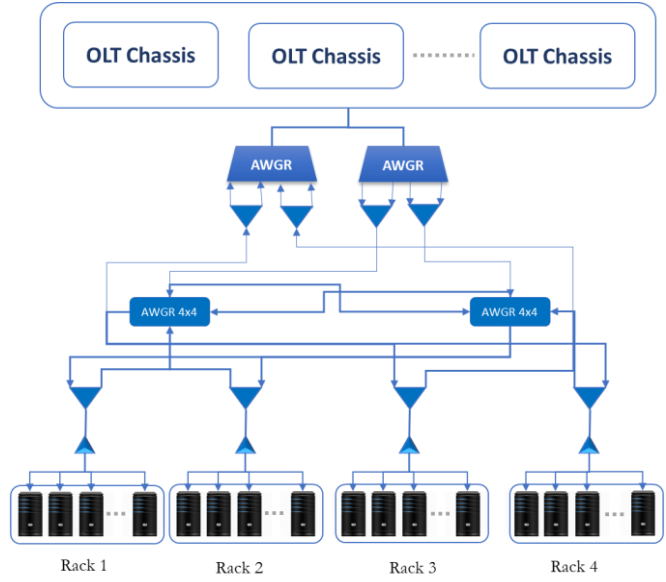

*Figure 1. AWGR Based Passive Optical Network Data Centre Architecture*

WDM/TDM-PON is implemented to provision the connection between the OLT and PON groups, as well as among the PON groups. This provides the ability to use several wavelengths sent over different time slots in various directions to connect the OLT, ONU and PON groups [21].

The optimisation objective is to maximise the number of connections, hence providing more time slots shared by a set of wavelengths. The results of the optimisation are compared to the results of the architecture when implementing a WDM AWGR based PON.

## 3. OPTIMISATION MODEL

The main purpose of using linear mathematical programming in designing this network is to determine the optimum number of wavelengths, time slots and routes required in order to achieve full interconnectivity within the architecture through mathematical linear expressions.

The parameters considered in this modelling are the number of nodes in the architecture which are servers, AWGRs and OLT/ONU. Other parameters are also considered such as: the links between nodes and their capacities, and demands between source and destination racks.

The architecture shown in Figure 1 is composed of four racks of servers. It uses four AWGRs to connect the racks to OLTs/ONUs, two of these AWGRs are 4x4 and the other two 1xN. The top layer is composed of several OLT chassis containing OLT/ONU. The data rate among servers is 10 Gbps, while on the OLT/ONU level it is 40 Gbps. All traffic is routed using 4 different wavelengths ($\lambda_1, \lambda_2, \lambda_3, \lambda_4$). Each wavelength has a capacity of 10 Gbps, divided into four time slots ($\tau_1, \tau_2, \tau_3, \tau_4$) each with a 2.5 Gbps rate. Demands on the network have been generated randomly using a uniform distribution between 1Gbps and 10Gpbs in steps of 2Gbps. Each rack is connected to one of the 4 ports of the lower tier AWGRs. The 4 different wavelengths received at the input port of the lower tier AWGRs will then be directed to the designated rack or to the higher tier AWGR to connect it to the OLT/ONU.

The objective of the optimisation is to minimise the number of granted resource blocks while ensuring that a certain set of demand volumes are satisfied. The objective function is written as:

Minimise:

$$\sum_{s \in P} \sum_{d \in P_{s \neq d}} \sum_{j \in W} \sum_{t \in T} \mu_{jt}^{sd}$$

where *s* is the source, *d* is the destination for a given demand, *j* is a single wavelength within the set *W* of all wavelengths, *and t* is a single time slot within the set *T* of all time slots; μ is a binary variable that takes the value '1' if wavelength *j* and time slot *t* are used between source and destination *'s,d'*, and takes the value '0' otherwise.

The model uses a flow conservation constraint to maintain demand routing between source and destination through intermediate nodes. A capacity control constraint ensures that the flow between a source and destination does not exceed the link capacity between them. A demand constraint is used to ensure that all demands are met. A wavelength and time slot constraint ensures that a single wavelength is used at all times between a certain source and destination, and that each destination receives a different wavelength on a given time slot from different sources. Each source sends to different destinations using different wavelengths and time slots taking into consideration that a given wavelength and time slot are not used more than once by a source and destination on the same link. The CPLEX solver was used to provide the solution and the model file is written using AMPL.

## 4. RESULTS

The results were generated by provisioning the wavelength routing and assignment of time slots for traffic flows within a PON cell among four different racks. The model is evaluated on 4 available wavelengths ($\lambda_1, \lambda_2, \lambda_3, \lambda_4$) as they are enough for the number of servers evaluated using only WDM. Each wavelength is divided into 4 time slots ($\tau_1, \tau_2, \tau_3, \tau_4$). Using the WDM/TDM PON gave the ability to use the same wavelengths for all communications within the PON cell, inter-rack and intra-rack communication. A single time slot within a wavelength is called a resource block. Resource blocks are granted as the lowest granularity, and multiple resource blocks are granted depending on the volume of each demand. The model described in Section 3 resulted in provisioning and optimising traffic routing using the available resources as shown in Figure. 2 and Figure. 3 showing the amount of resources allocated depending on the demands generated and traffic volume.

The results were generated in two scenarios; in scenario 1 as shown in figure 2, the traffic was generated using a uniform random distribution for WDM/TDM PON and all nodes were generating demands, while in scenario 2 as shown in figure 3 the traffic was generated using a uniform random distribution for WDM/TDM PON but the traffic is generated from a set of nodes while the rest are idle as shown in Figure 3. It is found that WDM/TDM PON is more efficient in utilising resources to maintain flow within the architecture in both scenarios as it shows in Figure 2 and Figure 3, the increase in demands and traffic resulted in increase in the number of resources needed to maintain the flow.

When employing WDM/TDM the architecture manages to relay traffic with up to 75% less resources than the case when only WDM is used. Resource savings are highest at low levels of demand and increase as the traffic volume per demand increases. This higher gain at lower demand values is due to the ability to use smaller granularities and therefore only a small portion of the overall activated wavelength is used.

Figure 3 shows how the resource allocation is affected when the number of demands with the same average traffic increases using both techniques, it is this shown that in the 'WDM only' architecture, resources are exhausted more rapidly than in the WDM/TDM architecture.

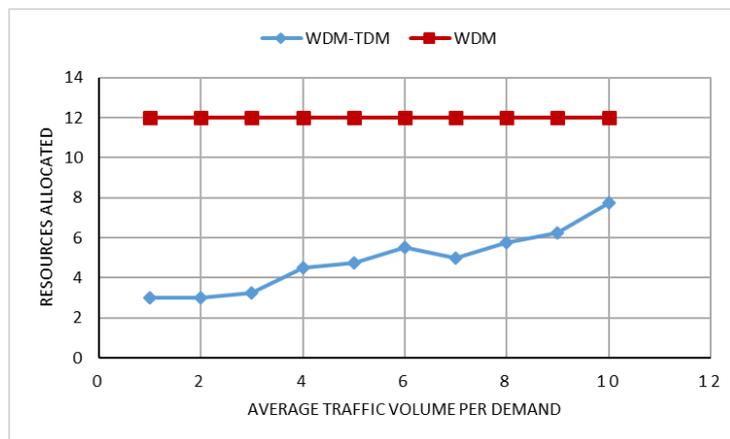

*Figure 2. Resources allocated vs. traffic volume*

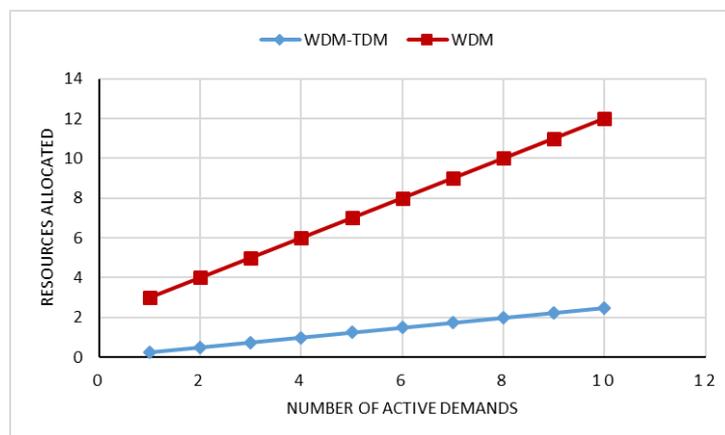

*Figure 3. Resources allocated vs. number of demands*

## 5. CONCLUSIONS

This paper introduced the use of WDM/TDM in an AWGR based passive optical network data centre architecture to realise more efficient resource utilisation by dividing the available blocks (wavelengths) into several time slots. A mathematical optimisation model was introduced for the architecture and the results were compared to the results of an optimisation of the same architecture while using WDM as a multiple access approach. The results showed that WDM/TDM is more efficient in resource utilisation than 'WDM only' architectures with resource savings reaching up to 75%.

### ACKNOWLEDGEMENTS

The authors would like to acknowledge funding from the Engineering and Physical Sciences Research Council (EPSRC), INTERNET (EP/H040536/1) and STAR (EP/K016873/1). All data are provided in full in the results section of this paper.